\documentclass[a4paper, 12pt]{article}

\usepackage{amsmath,amsthm,amssymb,amsfonts,amscd,latexsym}
\usepackage{graphicx,subfigure}
\newcommand{\Cov} {\mbox{$\rm{Cov}$\,}}
\newcommand{\Var} {\mbox{$\rm{Var}$\,}}
\newcommand{\Prob} {\mbox{$\rm{Pr}$\,}}

\title{Empirical distribution of $k$-word matches in biological sequences}

\begin{document}

\maketitle

\begin{center}
\textbf{Sylvain For\^et$^1$, Susan R.\ Wilson$^1$, Conrad J.\ Burden$^{1,2,3}$}
\end{center}

\small{
$^1$Centre for Bioinformation Science, Mathematical Sciences
Institute, The Australian National University, Canberra ACT 0200,
Australia\\
$^2$John Curtin School of Medical Research, The Australian National
University, Canberra ACT 0200, Australia\\
$^3$ Corresponding author, e-mail: Conrad.Burden@anu.edu.au}

\begin{abstract}
This study focuses on an alignment-free sequence comparison meth-od: the
number of words of length $k$ shared between two sequences, also known
as the $D_2$ statistic.
The advantages of the use of this statistic over alignment-based methods
are firstly that it does not assume that homologous segments are
contiguous, and secondly that the algorithm is computationally extremely
fast, the runtime being proportional to the size of the sequence under
scrutiny.
Existing applications of the $D_2$ statistic include the clustering of
related sequences in large EST databases such as the STACK database.
Such applications have typically relied on heuristics without any
statistical basis.
Rigorous statistical characterisations of the distribution of $D_2$ have subsequently been
undertaken, but have focussed on the distribution's asymptotic
behaviour, leaving the distribution of $D_2$ uncharacterised for most
practical cases.
The work presented here bridges these two worlds to give usable
approximations of the distribution of $D_2$ for ranges of
parameters most frequently encountered in the study of biological
sequences.
\end{abstract}

\section{Introduction}

\paragraph{}
The accelerating rate of accumulation of molecular sequences in public
databases has triggered the development of a number of sequence
comparison algorithms.
The most popular algorithms, such as FASTA, BLAST or BLAT, rely on
sequence alignment, and assume contiguity between homologous segments.
This assumption is, however, often broken in molecular sequences, due to
events such as transposition, unequal crossing over or alternative
splicing.
To address this issue, a number of alignment-free sequence comparison
methods have been developed.
Amongst them, the count of words of length $k$ letters matching between
two sequences, also known as the $D_2$ statistic, has found some
successful applications, due to its simplicity and its speed.
The algorithm to calculate the $D_2$ statistic between two sequences runs
as a linear function of the sequences' lengths, whereas alignment-based
sequence comparison methods typically have a worst case runtime
quadratic in the sequences' lengths.
The first applications of the $D_2$ statistic relied on heuristics to
decide whether sequences are significantly similar, but did not have any
statistical basis.

\paragraph{}
A rigorous examination of the distribution of $D_2$ led to the
characterisation of asymptotic distributions, but the behaviour of $D_2$ in
practical cases remains unknown.
In a previous study, we characterized $D_2$ optimal word sizes for a
range of sequence sizes.
The goal of the present study is to find approximations of the distribution of $D_2$
for word sizes close to optimal, and for the sequence
sizes most frequently encountered in molecular databases.

\section{Background}

The $D_2$ statistic is defined to be the number of exact word matches of
length $k$ between sequences ${\mathbf A} = (A_1, \ldots, A_m)$ and
${\mathbf B} = (B_1,\ldots, B_n)$, with $A_i$ and $B_j$ belonging to a
given alphabet $\cal A$.
For mathematical convenience we will impose periodic boundary conditions
on both sequences, that is, the letter in the first position in sequence
$\mathbf A$ is assumed to follow the letter in the $m$th position, and
the letter in the first position in sequence $\mathbf B$ is assumed to
follow the letter in the $n$th position.
For $k << m,n$ we do not expect our results to differ significantly from
the usual case of free boundary conditions.

\paragraph{}
Defining the indicator variables $Y_{(ij)}$ for a word match between the
$k$-word at position $i$ in A and the word at position $j$ in B by
\begin{equation}
Y_{(ij)} = \left\{
\begin{array}{ll}
1 & \mbox{if } (A_i, \ldots, A_{(i+k-1)\bmod m}) = (B_j, \ldots, B_{(j+k-1)\bmod n})\\
0 & \mbox{otherwise},
\end{array} \right.
\label{indicatorVariables}
\end{equation}
the $D_2$ statistic is given by
\begin{equation}
D_2 = \sum_{(i,j) \in I} Y_{(ij)},
\label{formulaForD2}
\end{equation}
where $I = \{(i, j) | 1 \le i \le m, 1 \le j \le n \}$.
For the case of free boundary conditions the index set is replaced by
$I = \{(i, j) | 1 \le i \le m - k + 1, 1 \le j \le n - k + 1 \}$.

\paragraph{}
We are interested in the distributional properties of $D_2$ under the
null hypothesis that $\mathbf A$ and $\mathbf B$ are Bernoulli texts,
meaning that each letter, $A_i$ or $B_j$, is independently and
identically (i.i.d.) distributed.
Let the probability of occurrence of letter $a \in {\cal A}$ be $f_a$,
and define
\begin{equation}
p_t = \sum_{a \in \mathcal{A}} {f_a}^t, \qquad t=1,2,\ldots.
\label{ptDefinition}
\end{equation}
The mean of $D_2$ is then~\cite{LHW02}:
\begin{equation}
E(D_2) = \sum_{(i,j) \in I} E(Y_{(ij)}) =
 mn\left(\sum_{a \in  \mathcal{A}} {f_a}^2\right)^k = mn{p_2}^k .         \label{meanOfD2}
\end{equation}
An exact value for the variance of $D_2$ has recently been given for the
case of free boundary conditions in \cite{KRS07}.  In Appendix I we
derive a similar formula for the variance for the algebraically
simpler case of periodic boundary conditions.

From here on, to simplify matters we set $m = n$.
Rigorous results exist for the limiting distribution of $D_2$ as $n
\rightarrow \infty$ in certain regimes.
For pairs of Bernoulli texts with non-uniform letter distributions, the
limiting distribution is compound Poisson in the regime
$k  > 2 \log_b n + \mbox{const.}$
\cite{LHW02}, and normal in the regime
$k < 1/2 \log_b n + \mbox{const}$
\cite{BKW07}.
Here $b = {p_2}^{-1}$.

\paragraph{}
In earlier numerical analyses~\cite{FKB06}, we tested the accuracy with
which $k$-word matches are able to measure the relatedness of
artificially evolved sequences.
Calculations of the optimum word size $k$ for a range of sequence
lengths $n$, showed that optimum word sizes generally fall between the two
parameter regimes for which the asymptotic behaviour of $D_2$ is known.
Our purpose here is to perform numerical experiments to fill in the gap
in the biologically relevant parameter regime between the asymptotically
normal and compound Poisson asymptotic behaviours, and to find accurate
and practical approximations to the distribution of $D_2$ in this parameter
region.
In particular, we are concerned with accurately reproducing the region
of the tail corresponding to classical significance levels (0.001\%, 0.01\%,
\ldots), both for the distribution of $D_2$, and for its extreme value
distribution that is used for determining p-values in database searches.

\section{Simulations of the empirical distribution}

\paragraph{}
The distribution of $D_2$ was simulated for a number of combinations of
sequence size $n$, word size $k$, alphabets $\cal A$ and sequence composition $f_a$.
For nucleic acid sequences, word sizes close to the optimal word size were chosen, based on
computation of the optimal word size of $D_2$ \cite{FKB06}.  We focused on sequence sizes typical of ESTs, whole genome shotgun sequencing trace pairs, CDSs, and mRNAs ($100 \le n \le 3200$ bases).  For protein amino acid sequences, the optimal word sizes and a letter composition
equal to the average of the proteins encoded by the human genome where
determined using the same method.  For protein sequences of length up to $n = 400$ the optimum word size was $k = 3$, and for longer sequences up to $n = 3200$ the optimum word size was $k=4$.  The sequence sizes for proteins ranged from small peptides to large
proteins (10 to 2560 residues).  Sequences were simulated with uniform and non-uniform letter
distributions.

\paragraph{}
For each combination of parameters, $N_\textrm{sample} = 10^6$ pairs of
Bernoulli text sequences were generated.
The extreme value distribution was simulated by taking the largest value
of 100 comparisons $N_\textrm{sample}$ times.
The code for the simulations was written in ANSI C and is available
from the author's website~\cite{prog}.

\section{Comparison between empirical and hypothesised distributions}

\paragraph{}
Previous studies of the $D_2$ statistic used Kolmogorov-Smirnov
tests~\cite{Conover99} to compare the empirical distribution of $D_2$
with its theoretical asymptotic distributions (normal or
compound-Poisson)~\cite{LHW02,BKW07}.
These studies, however, have been in error for the following reason.
Care must be taken when using the Kolmogorov-Smirnov test to pre-specify
the parameters of the distribution being compared.
If instead, parameters are estimated from the empirical distribution,
the p-values obtained will be overestimated (see Appendix II).
Given that these earlier studies generally pre-dated the discovery of an
analytic formula for the variance of $D_2$, they relied on means and
variances estimated from empirical samples, and therefore led to overly
optimistic claims of agreement between the distribution of the $D_2$
statistic for finite length sequences and its theoretical asymptotic
limit.

\paragraph{}
We have repeated Kolmogorov-Smirnov tests of our empirically generated
data, standardised with the analytically determined mean and variance of
$D_2$, against the standard normal distribution.
In general, we find p-values to be smaller than those reported in
earlier studies.
Similar results were obtained using the Shapiro-Wilk test, which tests
for normality but does not require prior knowledge of the mean or
variance.
More importantly, we find that the information provided by such
comparison is rather limited, as the p-value of the Kolmogorov-Smirnov
test decreases noticeably with the sample size $N_\textrm{sample}$,
since the true distribution of $D_2$ for finite sequence length $n$ never
exactly matches the hypothesised limiting distribution.
We conclude that this type of measure does not give a panacea for how well (or how badly) a given hypothesised distribution will
approximate the distribution of $D_2$.

\paragraph{}
Most practical uses of the $D_2$ statistic involve the calculation of a
p-value resulting from the comparison of two sequences or from the
comparison of a query sequence to a sequence database.
Our approach therefore is to compare a hypothesised distribution with an
empirically generated distribution of $D_2$ based on a direct comparison
of the p-values obtained with these two distributions.
If the p-values of a given hypothesised distribution agree well with those
of the empirical distribution, this hypothesised distribution could be
used to approximate the relevant tail of the real distribution of $D_2$.

\begin{figure}
	\centering
	\mbox{
		\subfigure[]{\includegraphics[width=6cm]{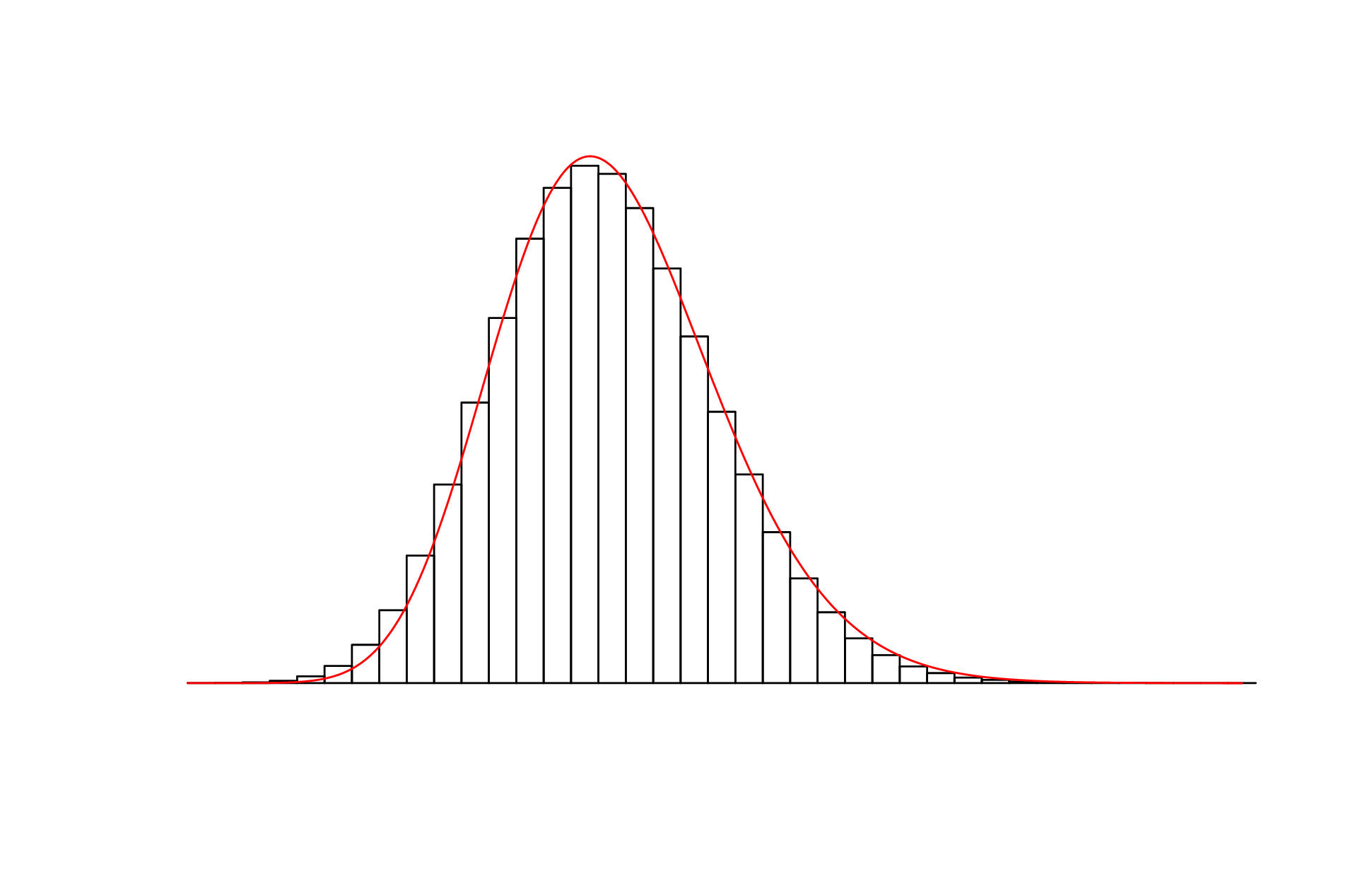}}
		\subfigure[]{\includegraphics[width=6cm]{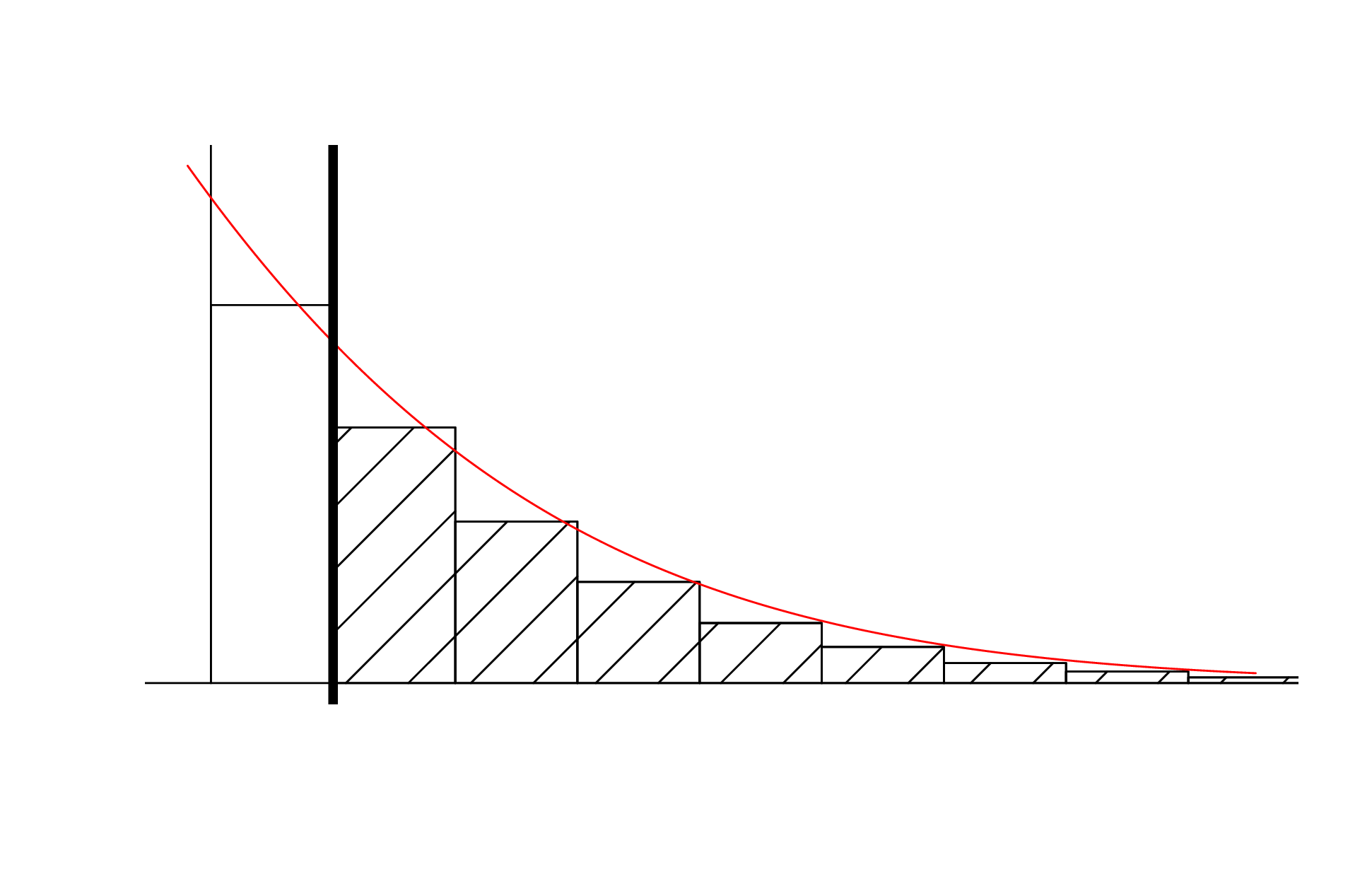}}
	}
	\caption{Distribution of $D_2$ for $n = 800$, $k = 7$.
	The histogram shows the empirical distribution, and the continuous curve is
	a hypothesised gamma distribution with mean given by
	Eq.~\ref{meanOfD2} and variance from the calculation in
	Appendix I.
	(a) Global view of the distribution.
	(b) Detail of the right hand tail of the distribution, the vertical
	bar shows the quantile $q_\textrm {hyp}$, the area under the curve is
	the corresponding level $p_\textrm {hyp}$, the hatched area is the
	empirical level $p_\textrm {emp}$.}
	\label{fig:empiricalPvalue}
\end{figure}

\paragraph{}
Suppose we wish to compare a postulated distribution function
$F_\textrm {hyp}$ with an empirically generated sample
$\{x_1, \ldots, x_{N_\textrm{sample}}\}$.
To evaluate how accurately p-values predicted by $F_\textrm {hyp}$
would approximate those of the true distribution of $D_2$, the quantiles
\begin{equation}
q_\textrm {hyp} = F_\textrm {hyp}^{-1}(1 - p_\textrm {hyp})
\end{equation}
are first calculated for to a number of p-values, $p_\textrm {hyp}$.
The frequency, in the simulated data, of the occurrences of $D_2$
greater than $q_\textrm {hyp}$ then provides an empirical p-value,
\begin{equation}
p_\textrm{emp} = \frac{\left| \{x_i: x_i \ge
q_{\textrm {hyp}}\} \right|}{N_\textrm{sample}}.
\end{equation}
This is compared to $p_\textrm {hyp}$:
\begin{equation}
\delta = \left| \log \left(\frac{p_\textrm{emp}}{p_\textrm {hyp}}\right) \right|.
\label{absoluteDiffEq}
\end{equation}
The comparisons focussed on p-values in the range of classical
significance levels ($p_\textrm {hyp}\in \{0.001\%, 0.01\%, \ldots\}$).
The theoretical distributions were parameterized using the exact values
of $D_2$'s mean and variance.
Zero values of $p_\textrm{emp}$ were replaced by
$1 / N_\textrm{sample}$.
The hypothesised distributions considered were the normal and gamma distributions.
The process is illustrated in Fig.~\ref{fig:empiricalPvalue}.

\paragraph{}
When doing database searches, a query sequence is compared to several
sequences, and the p-value of the best score of all these comparisons
needs to be estimated.
The relevant statistic in this case is the extreme value, that is, the
maximum of a number of i.i.d.\ random variables.
In addition to evaluating the tail of the distribution of $D_2$ itself, the tail of the empirical
extreme value distribution of $D_2$ was also compared to those of the the normal and the gamma distributions.
These two extreme value distributions belong to the Gumbel family and
can be easily computed (see Appendix III).

\section{Results}

\subsection{Approximating the distribution  of $D_2$}

\begin{figure}
	\centering
	\includegraphics[width=\textwidth]{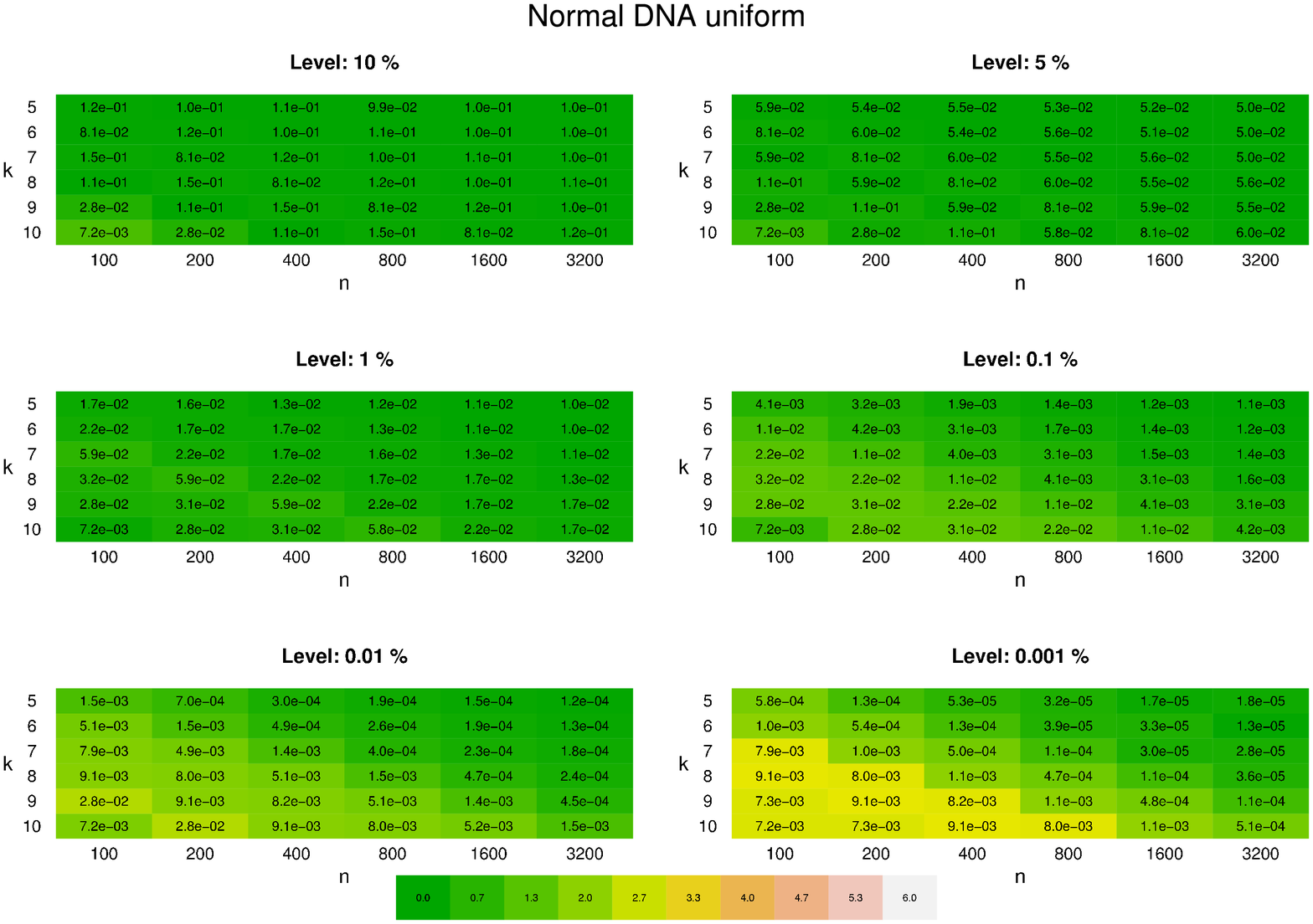}
	\caption{Normal distribution versus empirical distribution of $D_2$, DNA alphabet with uniform letter distribution.
	Each table compares the two distributions at a given level of the
	hypothesised distribution, for a number of combinations of
	sequence lengths $n$ and word sizes $k$.
	The value in each cell corresponds to the empirical level.
	The colour of each cell reflect the value of $\delta$, as
	introduced in Eq.~\ref{absoluteDiffEq}.}
	\label{fig:normalVsD2DNA}
\end{figure}

\begin{figure}
	\centering
	\includegraphics[width=\textwidth]{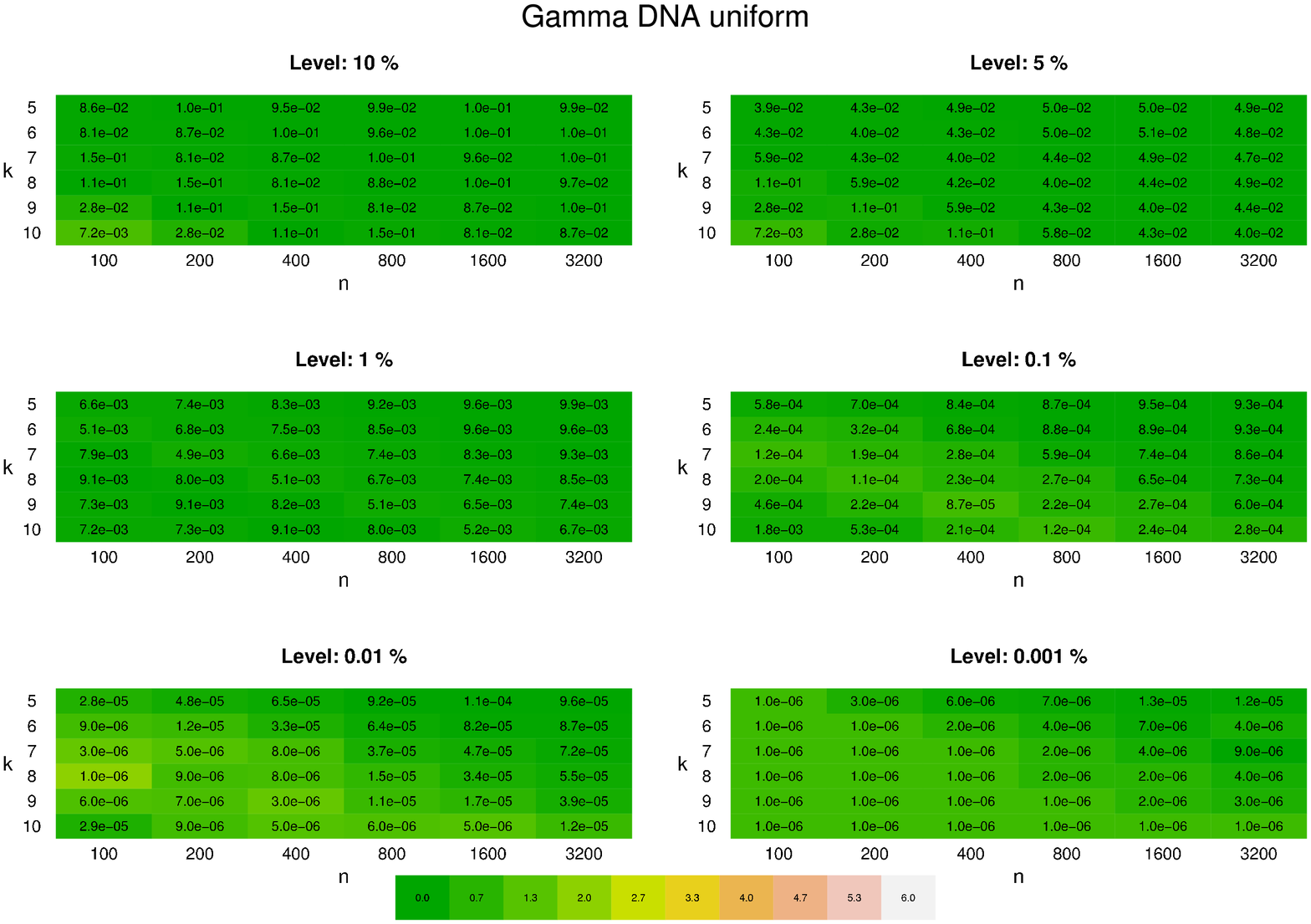}
	\caption{Gamma distribution versus empirical distribution of $D_2$,
	DNA alphabet with uniform letter distribution.
	See legend of figure~\ref{fig:normalVsD2DNA}.}
	\label{fig:gammaVsD2DNA}
\end{figure}

\paragraph{}
We first assessed the approximation of the distribution of $D_2$ with the
normal distribution.
Figure \ref{fig:normalVsD2DNA} shows the results of the comparison of the
p-values in the case of nucleic sequences with a uniform letter
distribution.
Similar results were obtained with non-uniform letter distributions.  
For sequences 1600 base pairs long or larger, the p-values from the hypothesised normal
distribution were very close to the empirical p-values.
For smaller sequences and large p-values (up to 1\%), the normal and
empirical p-values were of the same order of magnitude.
For smaller p-values, the hypothesised normal distribution greatly overestimated
the significance of $D_2$.
A few other distributions were compared to the simulated distribution of
$D_2$.
The gamma distribution, in particular, approximated the distribution of $D_2$
better than the normal distribution did.
In this case, the real p-values tended to be overestimated, and the
relative difference increased as the p-values decreased
(figure~\ref{fig:gammaVsD2DNA}).

\begin{figure}
	\centering
	\includegraphics[width=\textwidth]{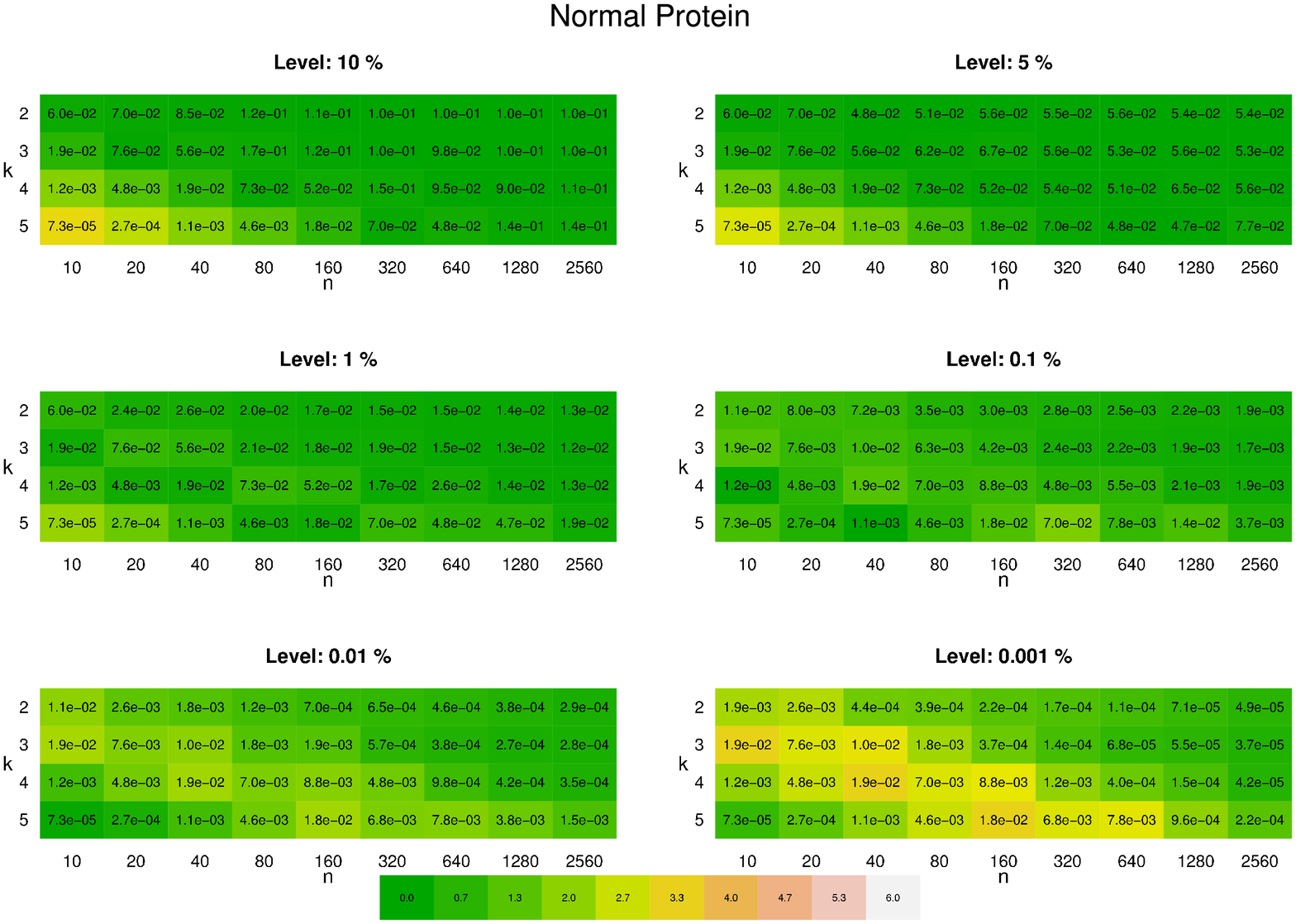}
	\caption{Normal distribution versus empirical distribution  of $D_2$, amino acid alphabet.
	See legend of figure~\ref{fig:normalVsD2DNA}.}
	\label{fig:normalVsD2AA}
\end{figure}

\begin{figure}
	\centering
	\includegraphics[width=\textwidth]{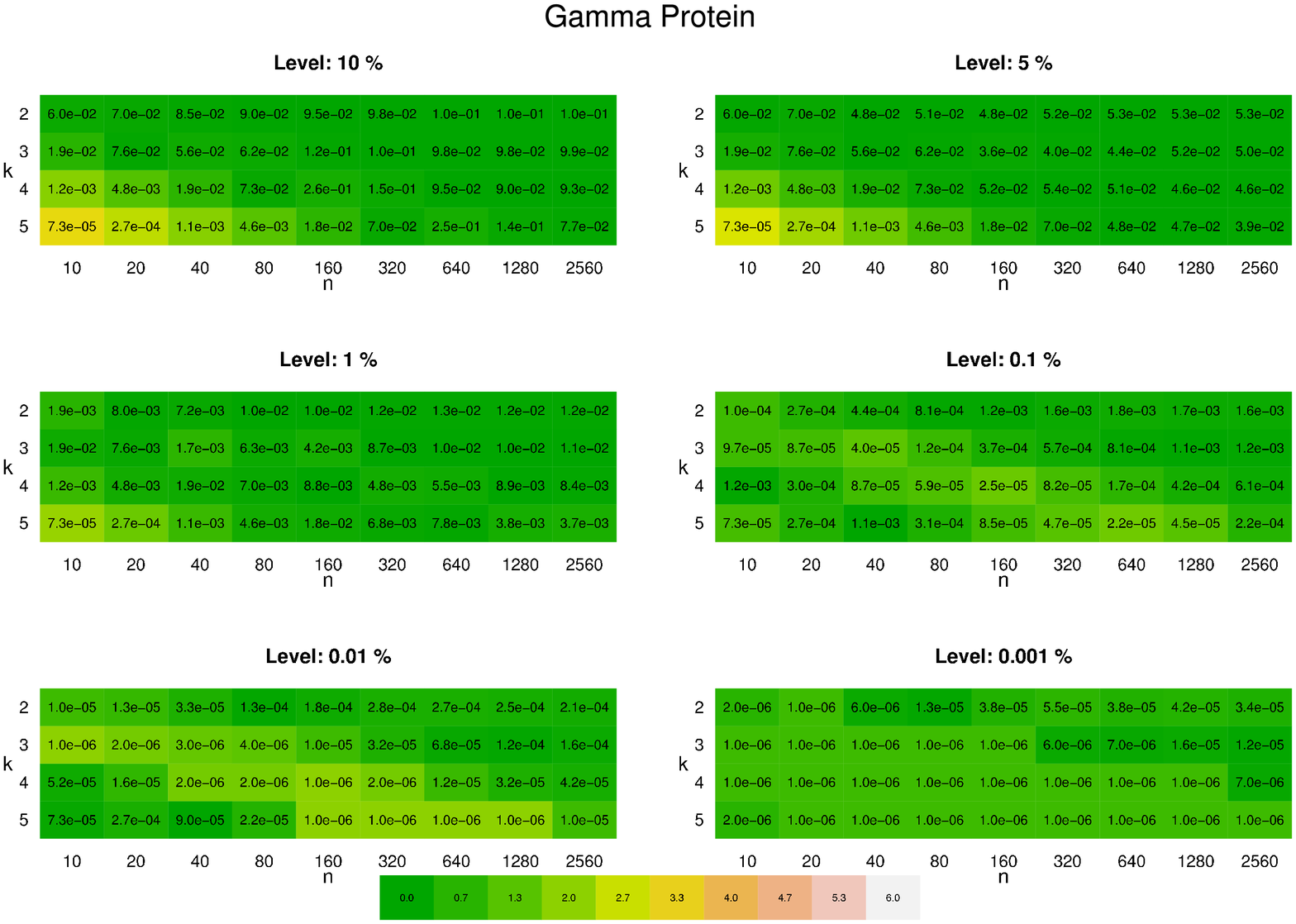}
	\caption{Gamma distribution versus empirical distribution of $D_2$, amino acid alphabet.
	See legend of figure~\ref{fig:normalVsD2DNA}.}
	\label{fig:gammaVsD2AA}
\end{figure}

\paragraph{}
The trends were identical for the amino acid alphabet (figure
\ref{fig:normalVsD2AA}): the
normal distribution approximates the p-values relatively well for large
sequences and moderate significance levels, but for shorter sequences
and further into the tail of the distribution, p-values were strongly overestimated.
The gamma distribution generally underestimated the p-values, but was
closer to the simulated distribution of $D_2$
(figure~\ref{fig:gammaVsD2AA}).

\subsection{Extreme value distribution}

\begin{figure}
	\centering
	\includegraphics[width=\textwidth]{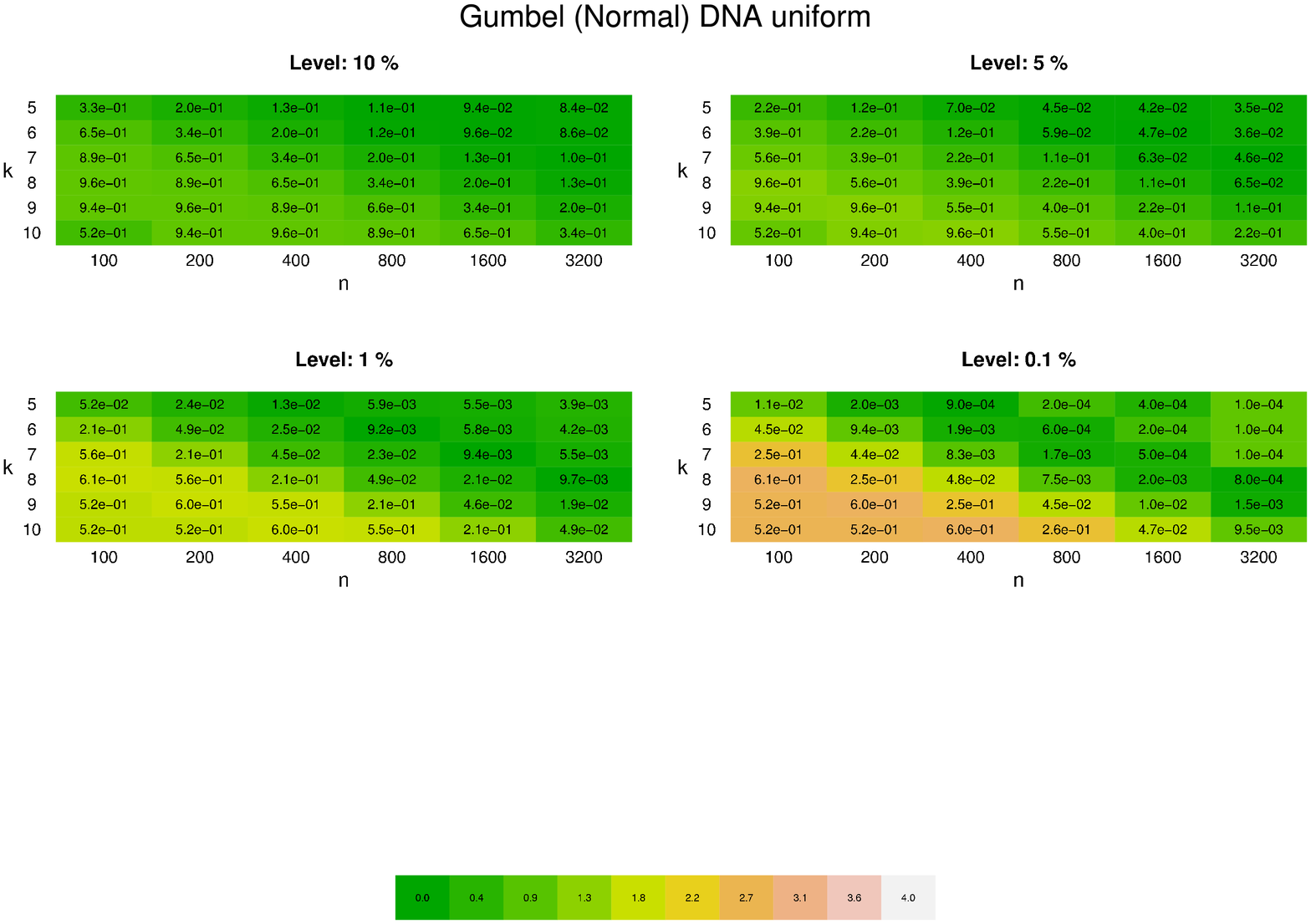}
	\includegraphics[width=\textwidth]{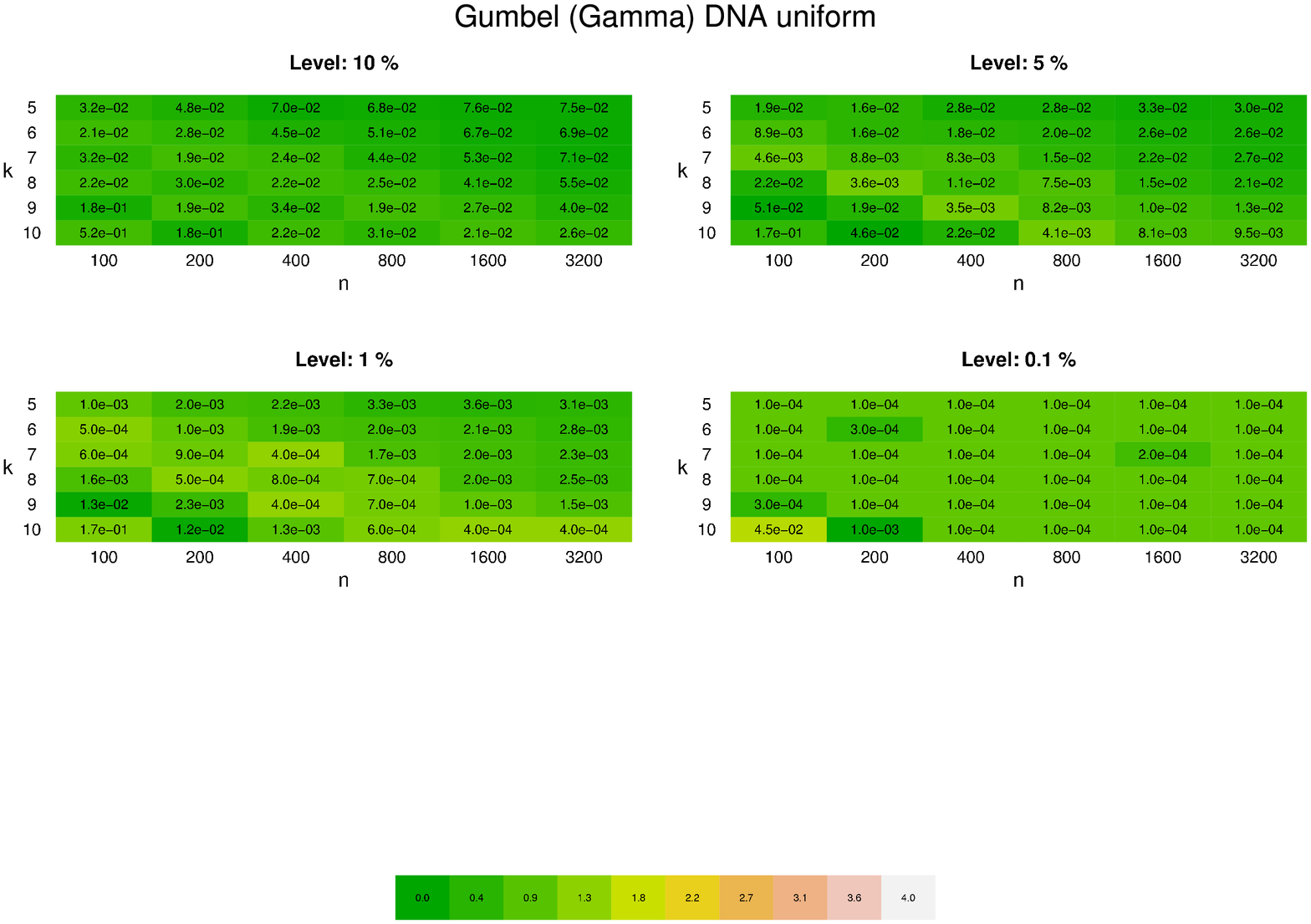}
	\caption{Extreme value of Normal and Gamma versus empirical extreme value
	of $D_2$, DNA alphabet with uniform letter distribution.
	See legend of figure~\ref{fig:normalVsD2DNA}.}
	\label{fig:D2EV}
\end{figure}

\paragraph{}
Figure~\ref{fig:D2EV} shows the results of comparison between the
extreme value distribution of $D_2$, and the extreme values of the gamma
and normal distributions in the case of a uniform nucleotide letter distribution for $p_\textrm {hyp}$ in the range $0.1\%$ to $10\%$.
The extreme value distribution of $D_2$ is generally better approximated by
the maximum of gamma distributions.
Since it was noted in the previous section that the relative difference
between the distribution of $D_2$ and the normal or the gamma distribution
increased as the p-values decreased, it is not surprising that the
approximations of the extreme value distribution of $D_2$ are not as good
as the approximations to the distribution of $D_2$.
The same trends were observed for nucleic sequences of non-uniform
letters and amino acid sequences.

\section{Discussion and Conclusions}

\paragraph{}
This study introduces practical approximations to the distribution of
the $D_2$ statistic and to the extreme value distribution of $D_2$.
For sequences of intermediate length (around 800 base pairs, close to
the average size of ESTs and sequencing traces) and for p-values between
5\% and 0.1\%, the Gamma distribution closely approximates the
distribution of $D_2$.
The Gamma distribution not only outperforms the normal distribution, but
unlike the latter, it slightly overestimates the p-values, and thus
would result in fewer false positives.

\paragraph{}
All the approximations presented here deteriorate as one moves further
to the right hand of the tail (for smaller p-values).
This is not, however, a major problem for any practical use of these
approximations, where very small p-values would just have an indicative
value.

\paragraph{}
Finally, our results show, that for longer sequences, such as genome
assembly contigs, the normal approximation itself would be appropriate,
even for very small p-values.

\section*{Appendices}

\subsection*{I. Calculation of $\Var D_2$}

Using Eq.~\ref{formulaForD2}, the variance of $D_2$ is
\begin{equation}
\Var(D_2) = \Var \left( \sum_{(i,j) \in I} Y_{(ij)} \right)
= \sum_{(i,j) \in I}  \Var(Y_{(ij)}) + \sum_{(i,j) \ne (i',j')} \Cov(Y_{(ij)}, Y_{(i'j')}).  \label{VarD2}
\end{equation}
To simplfy the notation from here on we set $u = (i,j)$, $v = (i',j')$.
The first term in Eq.~\ref{VarD2} depends only on
\begin{equation}
\Var(Y_u) = E(Y_u^2) - (E(Y_u))^2 = E(Y_u) - (E(Y_u))^2 = {p_2}^k - {p_2}^{2k},
\end{equation}
where $p_t$ is defined in Eq.~\ref{ptDefinition}.
Thus
\begin{equation}
\sum_{u \in I}  \Var(Y_u) = mn\left({p_2}^k - {p_2}^{2k}\right).
\label{varcontrib}
\end{equation}

To calculate the covariances in the second term of Eq.~\ref{VarD2}, it
is convenient to use the notation and terminology of~\cite{Wat95},
Chapter 11.
Let $J_u=\{v=(i',j') \, : \,  |i'-i| < k \, \mbox{ or }\, |j'-j| < k \}$
be the dependency neighbourhood of $Y_u$.
It can be decomposed into two parts, {\it accordion}
and {\it crabgrass},
$J_u = J_u^a \cup J_u^c$, where
$$
J_u^a = \{v=(i',j') \in J_u \, : \,  |i'-i| < k \, \mbox{ and }\, |j'-j| < k \} \} \quad
 \mbox{and} \quad J_u^c=J_u \setminus J_u^a.
$$
We compute the cross covariances, $\Cov (Y_u,Y_v)$, by looking at the
following cases.

\noindent
{\bf Case 1}: $v \not\in J_u$. In this case, $Y_u$ and $Y_v$ are
independent and hence $\Cov (Y_u, Y_v)=0$.

\noindent
{\bf Case 2}: $v \in J_u^c$.  Let $u=(i,j)$ and $v \in J_u^c$.
Consider first the subcase $v=(i+t, j')$,
where $|j-j'| \geq k$ and $ 0 \leq t \leq k - 1$.
Then
\begin{eqnarray}
E(Y_u Y_v) & = & \Prob(Y_u = 1, Y_v = 1) \nonumber \\
           & = & \sum_{(a_1, \ldots, a_{k+t}) \in \mathcal{A}}
	         (f_{a_1}\ldots f_{a_{k+t}})(f_{a_1}\ldots f_{a_k})(f_{a_{1+t}}\ldots f_{a_{k+t}}) \nonumber \\
           & = & \left(\sum_{a \in \mathcal{A}} {f_a}^2\right)^{2t}
                 \left(\sum_{a \in \mathcal{A}} {f_a}^3\right)^{k - t}  \nonumber \\
           & = & {p_2}^{2t} {p_3}^{k - t},
\end{eqnarray}
where we have used the fact that word matches occur simultaneously at
$u$ and $v$ if and only if the first $k$ letters of
$(A_i, \ldots, A_{i+ k + t -1})$
are repeated at $(B_j, \ldots, B_{j + k - 1})$ and the
final $k$ letters are repeated at $(B_{j'}, \ldots, B_{j' + k - 1})$.
This gives
\begin{equation}
\Cov(Y_u, Y_v) = E(Y_u Y_v) - E(Y_u)E(Y_v) = {p_2}^{2t} {p_3}^{k - t} - {p_2}^{2k}.
\end{equation}
Extending the argument to all $-k + 1 \le t \le k - 1$ gives
\begin{equation}
\Cov(Y_u, Y_v) = {p_2}^{2\left| t\right|} {p_3}^{k - \left| t\right|} - {p_2}^{2k}.
\end{equation}
By symmetry of the covariance function, the same result applies to the sub-case
$v=(i', j + t)$ where  $|i-i'| \geq k$ and $\left| t\right| \leq k - 1$.

The crabgrass contribution to the sum over covariance terms in
Eq.~\ref{VarD2} is then
\begin{eqnarray}
\lefteqn{ \sum_u \sum_{v \in J_u^c} \Cov(Y_u, Y_v)} & & \nonumber \\
     & = & \sum_u \left( \sum_{\{j':\left|j' - j\right| \ge k\} } +  \sum_{\{i':\left|i' - i\right| \ge k\} } \right)
           \sum_{t = -k+1}^{k-1} ({p_2}^{2\left| t\right|} {p_3}^{k - \left| t\right|} - {p_2}^{2k})  \nonumber \\
     & = & mn(m + n - 4k + 2) \left[ {p_3}^k+
           2 \sum_{t = 1}^{k - 1} {p_2}^{2t} {p_3}^{k - t}  - (2k - 1){p_2}^{2k}  \right] \nonumber  \\
     & = & mn(m + n - 4k + 2)
           \left[ {p_3}^k + 2{p_2}^2 p_3 \frac{{p_3}^{k - 1} - {p_2}^{2(k - 1)}} {p_3 - {p_2}^2}
           - (2k - 1) {p_2}^{2k} \right]  \nonumber \\
\label{case2contrib}
\end{eqnarray}

\begin{figure}
	\centering
	\includegraphics[height=8.1cm]{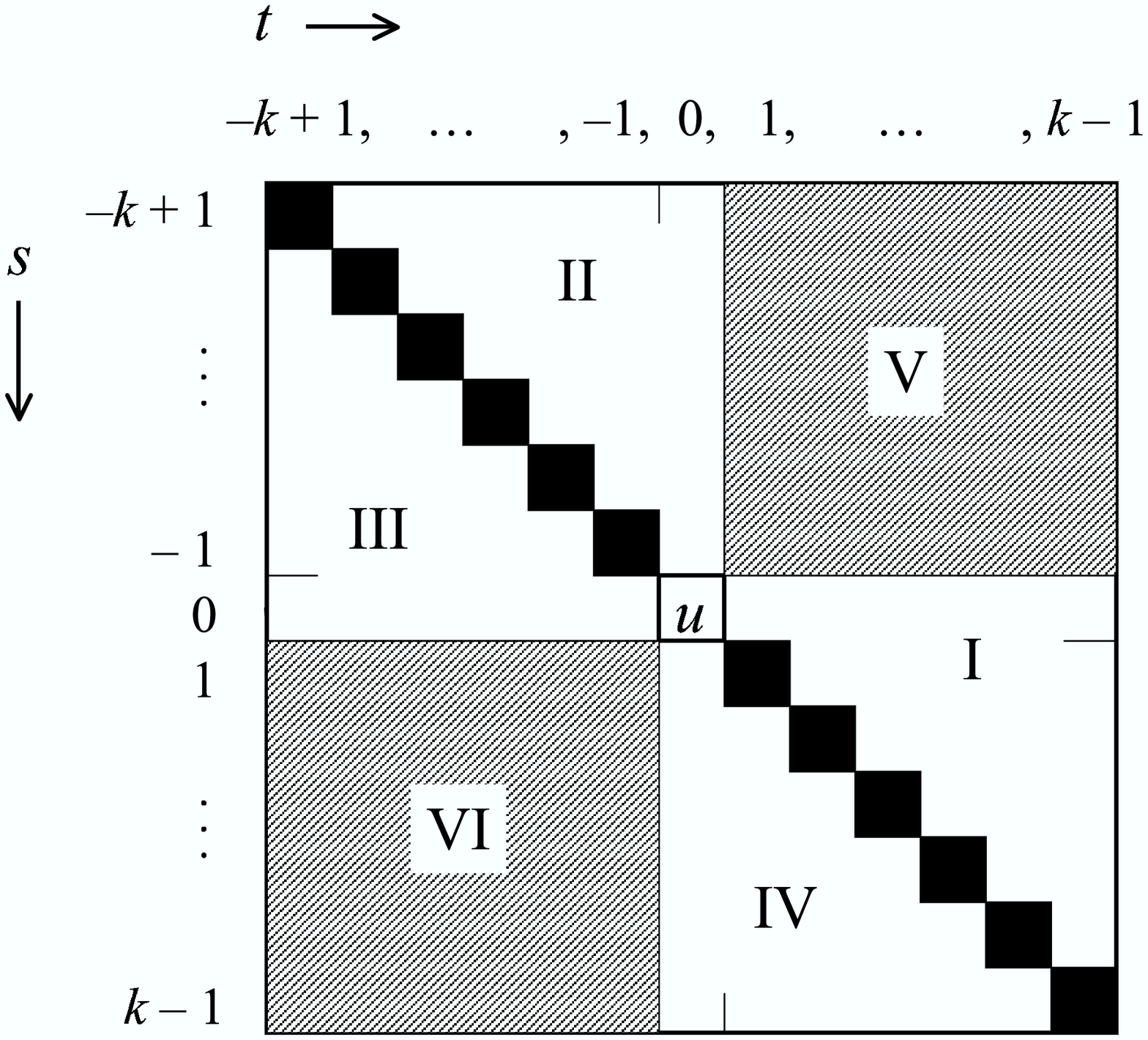}
	\caption{The main diagonal of $J_u^a$ referred to in Case 3 (black
	squares), and the sub-regions I to VI referred to in Cases 4 and 5.}
	\label{fig:accordion}
\end{figure}

\noindent
{\bf Case 3}: $v$ is on the {\it main diagonal} of $J_u^a$. That is,
$v = (i + t, j + t)$, where $-k < t < k$ and $t \neq 0$ (see Fig.~\ref{fig:accordion}).
In this case,
\begin{eqnarray}
E(Y_u Y_v) & = & \Pr (Y_u=1, Y_v=1) \nonumber \\
& = & \Pr (\mbox{a specific $(k + \left| t \right|)$-word match at the $(i,j)$ position} ) \nonumber \\
& = & \sum_{(a_1, \dots ,a_{k + \left| t \right|}) \in \mathcal{A}^{k + \left| t \right|}}
            {f_{a_1}}^2 \times \ldots \times {f_{a_{k + \left| t \right|}}}^2 \nonumber \\
& = & {p_2}^{k + \left| t \right|},
\end{eqnarray}
and
\begin{equation}
\Cov(Y_u, Y_v) = E(Y_u Y_v) - E(Y_u)E(Y_v) = {p_2}^{k + \left| t \right|} - {p_2}^{2k}.
\end{equation}
The contribution to the sum over covariance terms in Eq.~\ref{VarD2} from
Case 3 is then
\begin{eqnarray}
\lefteqn{ \sum_u \sum_{v \in {\rm main\, diagonal,}\, v \ne u} \Cov(Y_u, Y_v)} &   & \nonumber \\
                                                                               & = & 2mn \sum_{t = 1}^{k - 1} \left({p_2}^{t + k} - {p_2}^{2k}\right) \nonumber\\
                                                                               & = & 2mn \left[{p_2}^{k + 1} \frac{1 - {p_2}^{k - 1}}{1 - p_2} - (k - 1) {p_2}^{2k} \right].
\label{case3contrib}
\end{eqnarray}

\noindent
{\bf Case 4}: $v \in$ {\it one of the subregions I, II, III or IV of $J_u^a$ in Fig.~\ref{fig:accordion}}.
That is, $v = (i + t, j + s)$, where
\begin{itemize}
\item[I:] {$0 \le s < t \le k - 1$};
\item[II:] {$-k + 1 \le s < t \le 0$};
\item[III:] {$-k + 1 \le t < s \le 0$} or
\item[IV:] {$0 \le t < s \le k - 1$}.
\end{itemize}

\begin{figure}
	\centering
	\includegraphics[height=10cm]{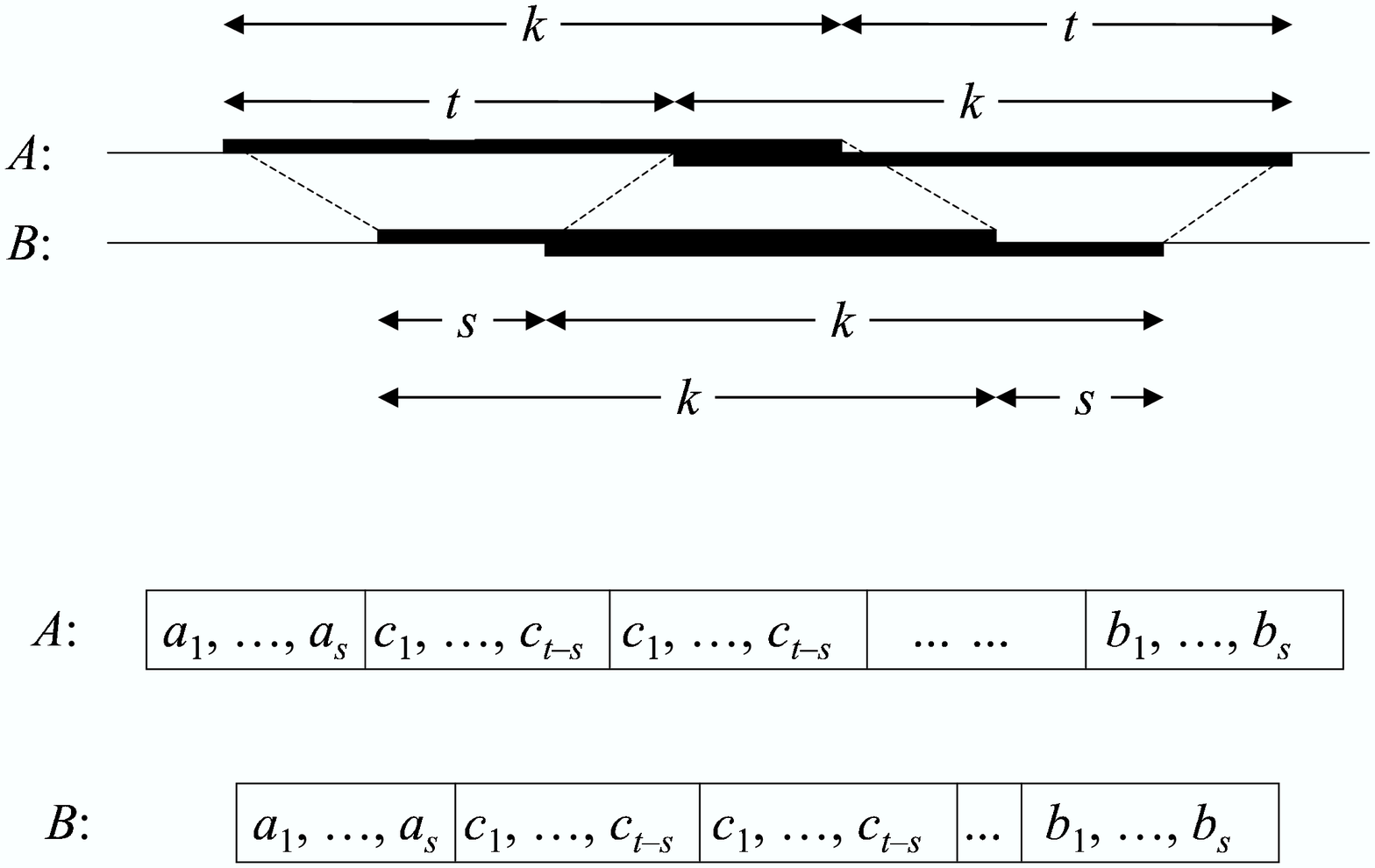}
	\caption{Word match configuration corresponding to Case 4(I).
	If the letters $a_1, \ldots, a_s$, $ b_1, \ldots, b_s$ and
	$ c_1, \ldots, c_{t - s}$ are specified, the remaining letters within
	the four words must be repeats of $c_1, \ldots, c_{t - s}$ as shown,
	the final repeat being truncated at the same point in both Sequence
	$A$ and Sequence $B$.}
	\label{fig:case4}
\end{figure}

Consider subregion I first.
The word matches corresponding to the event ``$Y_u = 1, Y_v = 1$'' are
illustrated in Fig.~\ref{fig:case4}.
For such a situation to occur, the $t + s$ letters $a_1, \ldots, a_s$,
$b_1, \ldots, b_s$ and $ c_1, \ldots, c_{t - s}$ can be specified
independently, and the remaining $2k$ letters within the four words must
be repeats of $c_1, \ldots, c_{t - s}$ as shown.
The sequence $ c_1, \ldots, c_{t - s}$ is repeated
$\nu = \lfloor(k - s)/(t - s)\rfloor$ complete times in sequence $B$ and
$\nu + 1$ complete times in sequence $A$, where $\lfloor \; \rfloor$
indicates the integer part.
At the right hand end of these repeats, the sequence
$ c_1, \ldots, c_\rho$ occurs once in Sequence $A$ and once in Sequence
$B$, where $\rho = (k - s)\bmod (t - s)$.

Then
\begin{eqnarray}
\lefteqn{E(Y_u Y_v) = \Pr (Y_u=1, Y_v=1)} & & \nonumber \\
& = & \sum_{(a_1, \ldots, a_s,  b_1, \ldots, b_s, c_1, \ldots, c_{t - s}) \in \mathcal{A}^{t + s}}
      {f_{a_1}}^2 \ldots {f_{a_s}}^2 {f_{c_1}}^{2\nu + 3} \ldots {f_{c_\rho}}^{2\nu + 3}
      \times     \nonumber \\
&  & \hspace{5cm}
     {f_{c_{\rho + 1}}}^{2\nu + 1} \ldots {f_{c_{t - s}}}^{2\nu + 1}  {f_{b_1}}^2 \ldots {f_{b_s}}^2
     \nonumber \\ \nonumber \\
& = & \left( \sum_{a \in \mathcal{A}} {f_a}^2 \right)^s
      \left( \sum_{c \in \mathcal{A}} {f_c}^{2\nu + 3} \right)^\rho
      \left( \sum_{c \in \mathcal{A}} {f_c}^{2\nu + 1} \right)^{t - s - \rho}
      \left( \sum_{b \in \mathcal{A}} {f_a}^2 \right)^s  \nonumber \\
& = & {p_2}^{2s} {p_{2\nu + 3}}^{\rho} {p_{2\nu + 1}}^{t - s - \rho} ,
\end{eqnarray}
and
\begin{equation}
\Cov(Y_u, Y_v) = E(Y_u Y_v) - E(Y_u)E(Y_v)
               = {p_2}^{2s} {p_{2\nu + 3}}^{\rho} {p_{2\nu + 1}}^{t - s - \rho} - {p_2}^{2k}.
\end{equation}

It is straightforward to check that similar results apply to subregions
II, III and IV, giving the contribution to the sum over covariances in
Eq.~\ref{VarD2} from Case 4 as
\begin{equation}
\sum_u \sum_{v \in R} \Cov(Y_u, Y_v)
= 4 nm \sum_{t = 1}^{k - 1} \sum_{s = 0}^{t - 1}
\left( {p_2}^{2s} {p_{2\nu + 3}}^{\rho} {p_{2\nu + 1}}^{t - s - \rho} - {p_2}^{2k} \right) ,
\label{case4contrib}
\end{equation}
where $R = \rm{I} \cup \rm{II} \cup \rm{III} \cup \rm{IV}$ is the union of the four subregions of Case 4 and 
\begin{equation}
\nu = \left\lfloor \frac{k - s}{t - s} \right\rfloor,  \qquad   \rho = (k - s) \bmod (t - s).
\end{equation}

\noindent
{\bf Case 5}: $v \in$ {\it one of the subregions V or VI of $J_u^a$ in Fig.~\ref{fig:accordion}}.
That is, $v = (i + t, j + s)$, where
\begin{itemize}
\item[V:] {$1 \le t \le k - 1$, $-k +1 \le s \le -1$};
\item[VI:] {$1 \le s \le k - 1$, $-k +1 \le t \le -1$}.
\end{itemize}

\begin{figure}
	\centering
	\includegraphics[height=10cm]{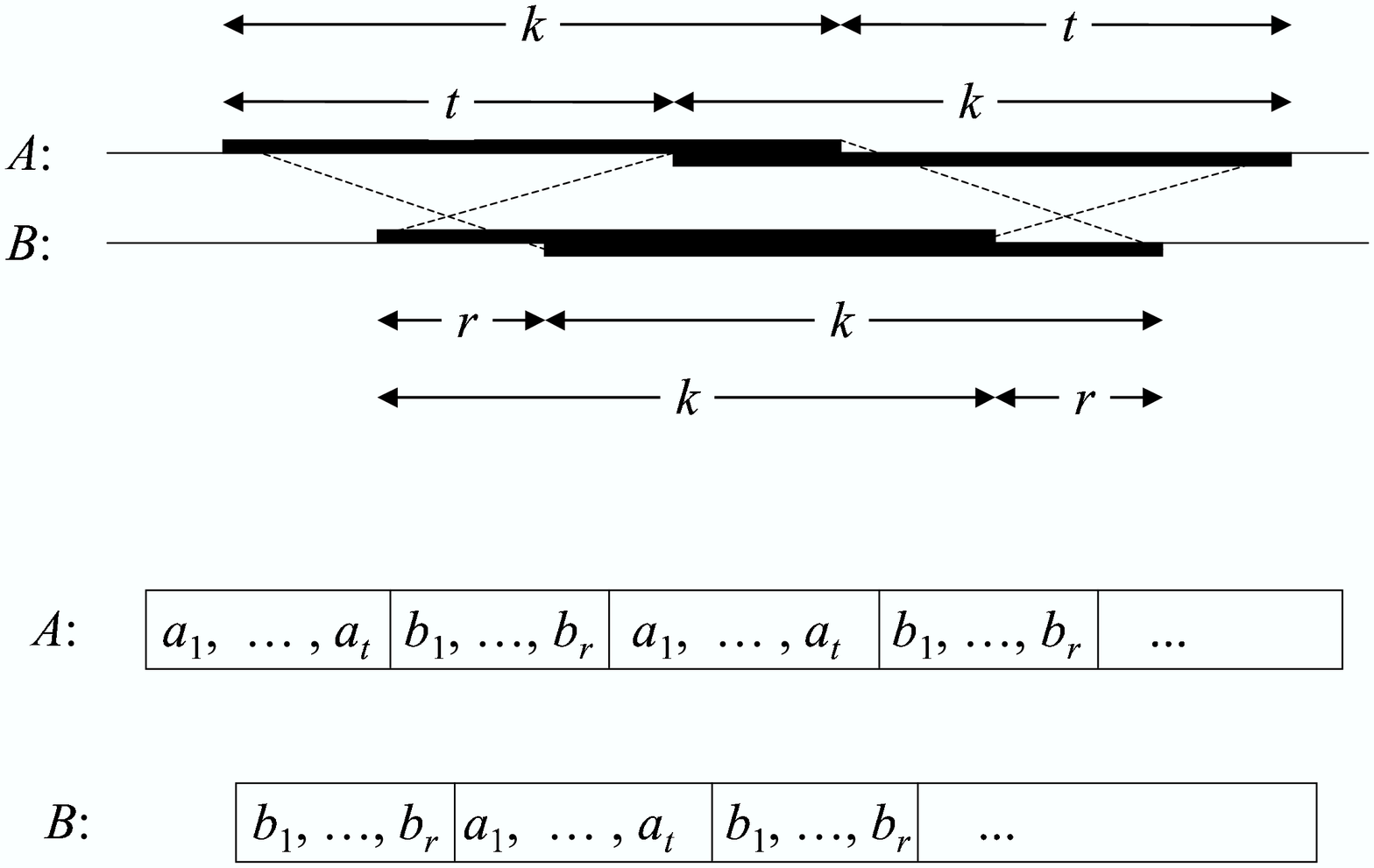}
	\caption{Word match configuration corresponding to Case 5(V).
	If the letters $a_1, \ldots, a_t$, $ b_1, \ldots, b_r$ are specified,
	the remaining letters within the four words must be repeats of
	$a_1, \ldots, b_r$ in Sequence A and $b_1, \ldots, a_t$ in sequence
	B, the final repeat being truncated at the $(k + t)$th or $(k + r)$th
	position respectively.}
	\label{fig:case5}
\end{figure}

Consider subregion V first.
The word matches corresponding to the event ``$Y_u = 1, Y_v = 1$'' are
illustrated in Fig.~\ref{fig:case5}.
Set $r = -s$.
For the event to occur, the affected block of length $t + k$ in Sequence
A must consist of repeats of a sequence
$(a_1, \ldots, a_t, b_1, \ldots, b_r)$ where
$a_1, \ldots, a_t$ and $b_1, \ldots, b_r$ are independently specified
letters.
The final repeat is truncated at the $(k + t)$th letter.
The affected block in sequence B must consist of repeats of the sequence
$(b_1, \ldots, b_r, a_1, \ldots, a_t)$, the final repeat being truncated
at the $(k + r)$th letter.

Let $l_i$, $i = 1, \ldots, t$ be the total number of times the letter
$a_i$ occurs and $m_j$, $j = 1, \ldots, r$ be the total number of times
the letter $b_j$ occurs in the two blocks in Fig.~\ref{fig:case5}.
By noting that the stretches of length $k$ not including the first $t$
letters of the A-block or not including the first $r$ letters of the
B-block each contain $\lfloor k/(r + t)\rfloor$ complete repeats of all
$s + t$ independent letters plus a final $k \bmod(r + t)$ remaining
letters at the right hand end, we arrive at
\begin{eqnarray}
l_i & = & 1 + 2\eta
            + \left\{ \begin{array}{ll} 1 & \mbox{if } i \le \zeta \\ 0 & \mbox{otherwise} \end{array} \right\}
            + \left\{ \begin{array}{ll} 1 & \mbox{if } i \le \zeta - r \\ 0 & \mbox{otherwise} \end{array} \right\}
                                                                                            \nonumber \\
m_j & = & 1 + 2\eta
            + \left\{ \begin{array}{ll} 1 & \mbox{if } j \le \zeta \\ 0 & \mbox{otherwise} \end{array} \right\}
            + \left\{ \begin{array}{ll} 1 & \mbox{if } j \le \zeta - t \\ 0 & \mbox{otherwise} \end{array} \right\},
\label{mndef}
\end{eqnarray}
where
\begin{equation}
\eta = \left\lfloor \frac{k}{r + t}\right\rfloor, \qquad \zeta = k \bmod(r + t).
\end{equation}
Then
\begin{eqnarray}
\lefteqn{E(Y_u Y_v) = \Pr (Y_u=1, Y_v=1)} & & \nonumber \\
& = & \sum_{(a_1, \ldots, a_t,  b_1, \ldots, b_r) \in \mathcal{A}^{t + r}}
      {f_{a_1}}^{l_1} \ldots {f_{a_t}}^{l_t} {f_{b_1}}^{m_1} \ldots {b_r}^{m_r}
      \nonumber \\
& = & \left(\sum_{a\in \mathcal{A}} {f_a}^{l_1}\right) \ldots \left(\sum_{a\in \mathcal{A}} {f_a}^{l_t}\right)
      \left(\sum_{b\in \mathcal{A}} {f_b}^{m_1}\right) \ldots \left(\sum_{b\in \mathcal{A}} {f_b}^{m_r}\right)
      \nonumber \\
& = & \left(\prod_{i = 1}^t p_{l_i} \right) \left(\prod_{j = 1}^r p_{m_j} \right),
\end{eqnarray}
and
\begin{equation}
\Cov(Y_u, Y_v) = E(Y_u Y_v) - E(Y_u)E(Y_v)
                = \left(\prod_{i = 1}^t p_{l_i} \right) \left(\prod_{j = 1}^r p_{m_j} \right) - {p_2}^{2k}.
\end{equation}

A similar result holds for subregion VI.  The contribution to the sum
over covariances in Eq.~\ref{VarD2} from Case 5 is then
\begin{equation}
\sum_u \sum_{v \in S} \Cov(Y_u, Y_v)
= 2 nm \sum_{r, t = 1}^{k - 1}
\left[ \left(\prod_{i = 1}^t p_{l_i} \right) \left(\prod_{j = 1}^r p_{m_j} \right) - {p_2}^{2k} \right] ,
\label{case5contrib}
\end{equation}
where $S = \rm{V} \cup \rm{VI}$ is the union of the two subregions of Case 5 and $l_i$ and $m_j$ are given by Eq.~\ref{mndef}.

Finally, by Eq.~\ref{VarD2}, the variance of $D_2$ is given by the sum
of the right hand sides of Eqs.~\ref{varcontrib}, \ref{case2contrib},
\ref{case3contrib}, \ref{case4contrib} and \ref{case5contrib}.

\subsection*{II. Consequences of failing to pre-specify parameters in the Kolmogorov-Smirnov test}

Given a random sample of observations
$X_1, X_2, \ldots, X_{N_\textrm{sample}}$, the Kolmogorov-Smirnov
test~\cite{Conover99} gives p-values for the null hypothesis that the
observations are associated with pre-specified distribution function
$F_\textrm{hyp}$.
The two-sided version of the test considered here uses as a test
statistic $\sup_i \left| F_\textrm{hyp}(X_i) - S(X_i) \right|$, where
$S$ is the empirical cumulative distribution function based on the
observations.
Under the null hypothesis the p-values obtained are uniformly
distributed on the interval $[0, 1]$.

\begin{figure}
	\centering
	\includegraphics[height=12cm]{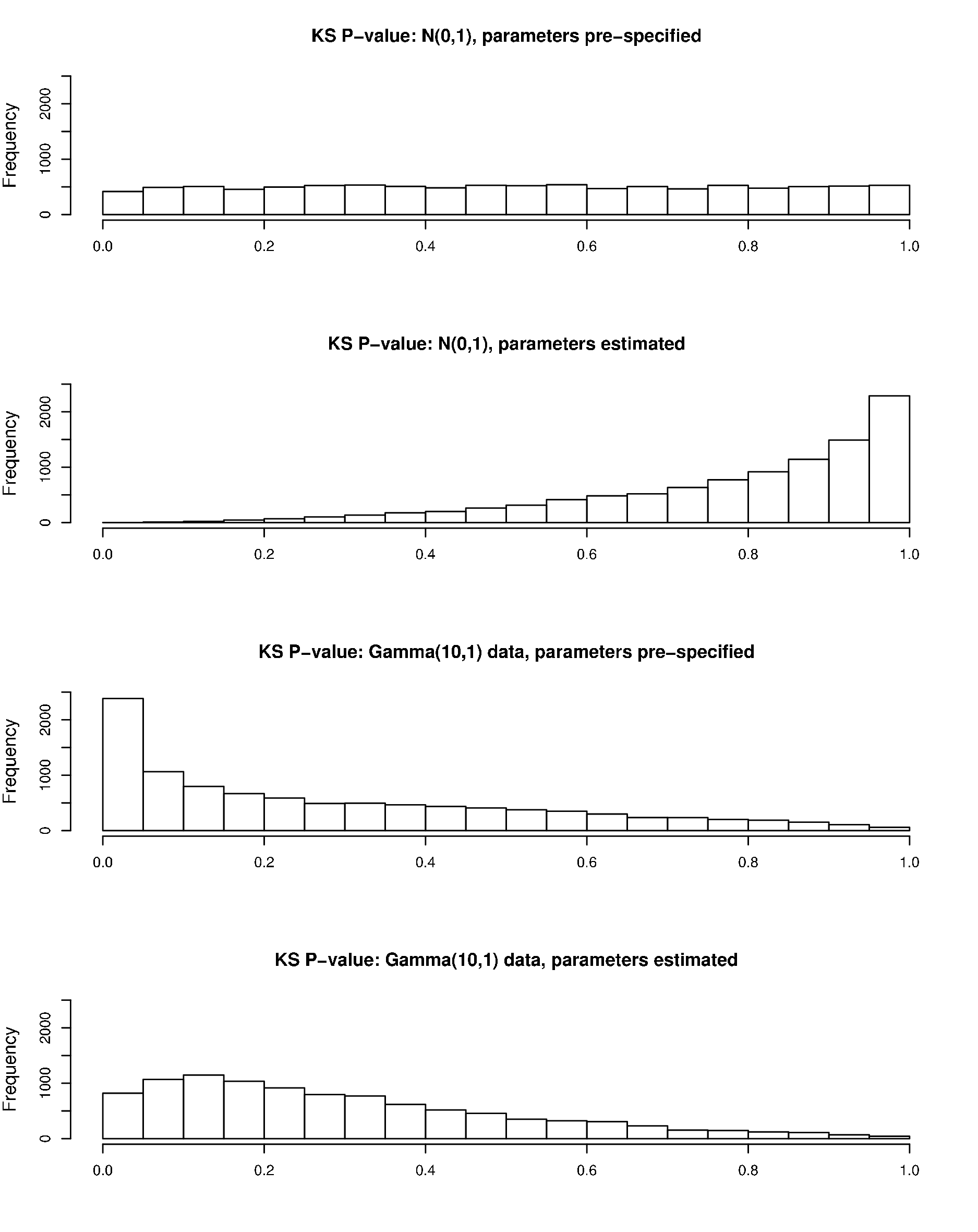}
	\caption{Histograms of p-values obtained from the Kolmogorov-Smirnov
	test applied to artificially generated data tested against a normal
	distribution.
	From top to bottom the plots are
	(i) standard normal data, parameters of the hypothesised distribution
	pre-specified;
	(ii) standard normal data, parameters of the hypothesised distribution
	estimated from the data;
	(iii) gamma distributed data with mean 10 and variance 1, parameters
	of the hypothesised distribution pre-specified; and
	(iv) gamma distributed data with mean 10 and variance 1, parameters
	of the hypothesised distribution estimated from the data.}
	\label{fig:KS_normal}
\end{figure}

Importantly, if the hypothesised distribution $F_\textrm{hyp}$ is not
fully pre-specified, but relies on estimates from the sample, the
reported p-values will not be uniformly distributed under the null
hypothesis.
To illustrate this, we have generated a set of 10,000 independent
samples of $N_\textrm{sample} = 2500$ random numbers from a standard
normal distribution, and applied the two-sided Kolmogorov-Smirnov test
to each sample using the R function {\tt ks.test}.
Histograms of the p-values obtained are shown in
Fig.~\ref{fig:KS_normal}.
In the first plot each sample was tested against the standard normal $N(0,1)$, whereas in the
second plot each sample was tested against  a normal distribution
whose mean and variance was estimated from the sample.  
We see that in this situation, where the null hypothesis is true, but
the Kolmogorov-Smirnov test is applied incorrectly, p-values are skewed
heavily towards 1.

In a second test to see whether incorrect use of the Kolmogorov-Smirnov
test can lead to an overly optimistic indication of agreement with a
hypothesised distribution, we generated a set of 10,000 independent
samples of $N_\textrm{sample} = 2500$ random numbers from a Gamma
distribution with mean 10 and variance 1.
This distribution is close to, but not identical with, the normal
distribution $N(10,1)$.
The third and fourth histograms in Fig.~\ref{fig:KS_normal} are of
p-values obtained from application of the Kolmogorov-Smirnov test
against a pre-specified $N(10,1)$, and against a normal distribution with mean and variance estimated from the sample respectively.  The third plot is an indication the distribution of p-values that will result if the Kolmogorov-Smirnov test for normality is applied correctly to this non-normal data.  
Again we see that p-values are overestimated in the fourth plot when the test is used
incorrectly.

\subsection*{III. Limiting distribution of the maximum of $N$ i.i.d.\ random variables}

We are interested in the limiting distribution for $N$ large of the
random variable
\begin{equation}
X_{\rm max} = \max_i X_i,
\end{equation}
where $X_i, i = 1, \ldots, N$ are i.i.d.\ random variables with common
density function $f_X$ and distribution function
\begin{equation}
F_X(x) = \int_{-\infty}^x  f_X(\xi) d\xi.
\end{equation}
The general theory of extreme value distributions is given in the book
by Gumbel~\cite{Gumbel58}, Chapter~5.2.
For distributions of ``type I'', which includes the normal and gamma
distributions, the distribution function of $X_{\rm max}$, namely
$(F_X(x))^N$, asymptotes to the double exponential distribution function
\begin{equation}
G(x) = \exp(-e^{-y}),
\end{equation}
where the {\em reduced largest value} is defined as
\begin{equation}
y = \alpha_N(x - u_N).
\end{equation}
Here $u_N$, called the {\em characteristic largest value}, is determined
by the condition that in $N$ observations of $X$, the expected number of
values greater than or equal to $u_N$ is unity.
It is the solution to the equation
\begin{equation}
F_X(u_N) = 1 - \frac{1}{N},
\end{equation}
and for the case of the normal and gamma distribution is easily found
using the R function {\tt qnorm()} and {\tt qgamma()} respectively.
The parameter $\alpha_N$ is called the {\em extremal intensity function}
and is given by
\begin{equation}
\alpha_N = Nf_X(u_N).
\end{equation}

\section*{Acknowledgement}
We thank Professors Alan Welsh and Chris Field for inspiring
discussions.
This work was funded in part by ARC Discovery Grant DP0559260.

\section*{Vitae}

\subsection*{Sylvain For\^et}

\paragraph{}
Sylvain For\^et graduated in Invertebrates Physiology and Computer Science at
the INAP-G and Universit\'e Pierre et Marie Currie in Paris, France in
1998.
He received his PhD in molecular biology and biochemistry from the
Australian National University in Canberra, Australia in 2007.

\subsection*{Susan R.\ Wilson}

\paragraph{}
Susan Wilson obtained her B.Sc.\ from the University of Sydney in 1968, and
her Ph.D. from the Australian National University in 1972. She is an
elected member of the International Statistical Institute, elected Fellow
of the American Statistical Association and of the Institute of
Mathematical Statistics. She has been President, International Biometric
Society.

\subsection*{Conrad J.\ Burden}

\paragraph{}
Conrad J.\ Burden received his B.Sc. in applied mathematics from
the University of Queensland in 1978, his Ph.D.\ in theoretical physics
from the Australian National University in 1983, and is a Fellow of the
Australian Institute of Physics.
For the first 16 years of his academic career his research interests
centred on subatomic particle physics and quantum field theory.
After a brief sojourn in the IT industry he made the transition to
bioinformation science in 2003.
He is currently a Fellow in the Centre for Bioinformation Science at the
Australian National University where his research interests include
modelling of oligonucleotide microarrays,  alignment free sequence
comparison methods, gene regulation and protein structure.

\end{document}